\documentclass[12pt]{article}
\usepackage{graphicx}
\catcode`\@=11
\def\numberbysection{\@addtoreset{equation}{section}
\def\theequation{\thesection.\arabic{equation}}}
\numberbysection

\newcommand{\beq}{\begin{equation}}
\newcommand{\beqa}{\begin{eqnarray}}
\newcommand{\eeq}{\end{equation}}
\newcommand{\eeqa}{\end{eqnarray}}
\newcommand{\abs}[1]{\vert#1\vert}
\newcommand{\absln}[1]{\vert\!\ln#1\vert}
\renewcommand{\d}{{\rm d}}
\newcommand{\dy}{{\rm dyn}}
\newcommand{\e}{{\rm e}}
\newcommand{\eps}{\varepsilon}

\renewcommand{\flat}{{\rm flat}}
\renewcommand{\frac}[2]{\displaystyle{\displaystyle#1\over\displaystyle#2}}
\newcommand{\h}{h}

\newcommand{\lam}{\lambda}
\newcommand{\lra}{\Longrightarrow}
\renewcommand{\max}{{\rm max}}
\renewcommand{\min}{{\rm min}}
\newcommand{\mean}[1]{\langle#1\rangle}
\newcommand{\prob}{\mathop{\rm Prob}\nolimits}
\newcommand{\s}{\sigma}
\newcommand{\st}{{\rm stat}}
\newcommand{\sign}{\mathop{\rm sign}\nolimits}
\newcommand{\xidy}{\xi_\dy}
\newcommand{\A}{{\cal A}}
\newcommand{\C}{{\cal C}}
\newcommand{\F}{{\cal F}}
\newcommand{\Frac}{\mathop{\rm Frac}\nolimits}
\newcommand{\Int}{\mathop{\rm Int}\nolimits}
\newcommand{\N}{{\cal N}}
\renewcommand{\P}{{\cal P}}
\newcommand{\T}{\Gamma}

\begin{document}
\centerline{\Large\bf A column of grains in the jamming limit:}
\vspace{.3cm}
\centerline{\Large\bf glassy dynamics in the compaction process}
\vspace{1cm}
\centerline{\large J.M.~Luck$^{a,}$\footnote{luck@spht.saclay.cea.fr}
and Anita Mehta$^{b,}$\footnote{anita@boson.bose.res.in}}
\vspace{1cm}
\noindent $^a$Service de Physique Th\'eorique\footnote{URA 2306 of CNRS},
CEA Saclay, 91191 Gif-sur-Yvette cedex, France

\noindent $^b$S.N.~Bose National Centre for Basic Sciences, Block JD,
Sector 3, Salt Lake, Calcutta 700098, India
\vspace{1cm}
\begin{abstract}
We investigate a stochastic model
describing a column of grains in the jamming limit,
in the presence of a low vibrational intensity.
The key control parameter of the model, $\eps$,
is a representation of granular shape, related to the reduced void space.
Regularity and irregularity in grain shapes,
respectively corresponding to rational and irrational values of $\eps$,
are shown to be centrally important in determining
the statics and dynamics of the compaction process.
\end{abstract}

\vfill
\noindent P.A.C.S.: 45.70.--n, 45.70.Cc, 45.70.Mg

\newpage
\setcounter{footnote}{0}
\section{Introduction}

The study of slow dynamics in the jamming limit unifies
the fields of granular compaction~\cite{sam,pgg} and glasses~\cite{glassyref}.
Key features of this involve frustration and hysteresis, among
other complex phenomena~\cite{spinglass},
with the concomitant difficulty of modelling them in simple and physical ways.
We present in the following a model of remarkable simplicity,
which is nevertheless
able to capture to a large extent the complex consequences of non-trivial
interactions, {\it even in one dimension}.
Issues that are probed include the effects of orientation,
and thus shape, on packing in the jamming limit.
`Irregular' and `regular' shapes of units (for example, grains)
will be seen to have
rather different consequences for compaction behaviour,
when they are subjected to zero- and low-temperature dynamics.

The present model is an extension, with interactions,
of an earlier model of non-interacting grains, presented in~\cite{usepl}.
For clarity, we summarise in what follows the commonality and differences
between the two.
The previous model was two-dimensional.
Each lattice site was occupied by an ordered grain ($+$),
a disordered grain ($-$), or a hole ($0$).
It interpolated between
a fluidised regime, where there were many holes on the lattice,
and a jammed regime, where there were no holes anywhere on the lattice.
The restricted dynamics in the jammed regime forbade
migration of grains anywhere on the two-dimensional lattice,
and in particular between columns.
The result was
expressible in terms of a column model of noninteracting grains,
with a (trivial) ground state of completely ordered ($+$) grains.
This, surprisingly, nevertheless exhibited some features usually associated
with glassiness, such as slow dynamics and aging~\cite{usepl}.

Clearly, disordered systems such as glasses or jammed granular media
do not have ground states that are crystalline; equally,
their attempts to reach their ground states are mediated
by complex long-range interactions.
The present model, already introduced in~\cite{usproc}
and investigated to some extent in~\cite{usletter},
represents an effort to make the jammed limit
of the earlier model more realistic, by the inclusion of interactions.
The column contains no holes.
Each grain is either ordered (represented by a ($+$) Ising spin)
or disordered (represented by a ($-$) spin).
However, and differently from the previous model,
we take into account the effect of voids
(holes that {\it partially} occupy a lattice site)
which are associated with each disordered orientation of a grain.
Thus, while each ordered grain fully occupies one unit of space,
each disordered grain occupies $\eps$ units of space,
so that $(1-\eps$) is a measure of the trapped void space.
The net volume occupied by a disordered grain,
or the corresponding void space, will depend on its shape;
we see that {\it $\eps$ is thus a simple representation of granular shape}.
Also, and differently from the earlier model, the process of compaction is now
no longer a simple relaxation into a completely ordered state:
a given distribution of ($+$) and ($-$) grains, as in nature, responds
to externally imposed dynamics, by minimising void space {\it locally},
in the presence of disorder.
The ground states so obtained resemble much more
the random close-packed state found in granular systems~\cite{bernal}
than the rather unrealistic crystalline state (all grains ordered)
obtained before.

\section{The model}

Our model is a fully directed model of interacting grains,
where causality induces a directionality both in time and in space,
as the orientation of a given grain only influences the grains
{\it below} it, and at {\it later} times.
Grains occupy all $N$ sites of a column.
As said above, we assume that they only assume two orientational states.
We set $\s_n=+1$ (resp.~$\s_n=-1$) if grain $n$ is ordered (resp.~disordered).
A configuration of the system is uniquely defined
by the orientation variables $\{\s_n\}$.

It is known~\cite{sid, gcbam} that the response
of jammed granular media to low-amplitude vibration
is compacting; that is, grains rearrange themselves
to locally maximise their packing fraction or, equivalently,
minimise their void space.
We model this by a local stepwise compacting dynamics: that is, a given grain
orients itself to minimise void space {\it locally},
given the orientations of grains above itself.

To be more specific,
in the presence of a dimensionless vibration intensity $\T$,
we consider a stochastic dynamics, defined by the orientation-flipping rates
\beq
\left\{\matrix{
w_n(+\to-)=\exp(-(\lam_n+\h_n)/\T),\hfill\cr
w_n(-\to+)=\exp(-(\lam_n-\h_n)/\T).\hfill
}\right.
\label{rates}
\eeq
In these expressions, $\h_n$ and $\lam_n$ are, respectively,
the local ordering field and the activation energy felt by grain~$n$.
These quantities are assumed to
only depend on the orientations of grains {\it above} grain~$n$.

We make the further simplifying assumption that the activation energy~$\lam_n$
does not depend on grain orientations at all.
We write
\beq
\lam_n=n\T/\xidy,
\eeq
where $\xidy$ is defined to be the {\it dynamical length}~[see~(\ref{p})].
Roughly speaking, $\xidy$ is the depth of the nonequilibrium boundary layer:
grains which are well within this length can order
relatively freely in response to surface events,
while grains much deeper relax only logarithmically in time~\cite{usepl}.

The only dependence of the dynamics
on orientations is via the ordering field $\h_n$,
which determines the orientational response of grain~$n$
to orientations of grains above it.
We choose to write the simple, linear formula
\beq
h_n=\eps m^-_n-m^+_n,
\label{ydef}
\eeq
where $m^+_n$ and $m^-_n$ are the numbers of ordered
and disordered grains above grain~$n$:
\beq
m^+_n=\frac{1}{2}\sum_{k=1}^{n-1}(1+\s_k),\qquad
m^-_n=\frac{1}{2}\sum_{k=1}^{n-1}(1-\s_k).
\eeq

In spite of its simplicity,
the dynamics defined by~(\ref{rates})--(\ref{ydef})
can be shown to capture the compaction mechanism sketched in the Introduction,
namely a local minimisation of the {\it excess void space}~\cite{brownrichards}.
In particular, a transition from the ordered to the disordered state
for grain~$n$ is
{\it hindered} by the number of voids that are already above it.
The whole picture will become clearer with the example of $\eps=1/2$,
discussed in section~3.2.

In order to perform numerical Monte-Carlo simulations
we will need a discrete-time formulation of the above rules.
The flipping rates $w_n$ become flipping probabilities
\beq
\left\{\matrix{
p_n(+\to-)=\frac{P_n}{1+\exp(2\h_n/\T)},\hfill\cr\cr
p_n(-\to+)=\frac{P_n}{1+\exp(-2\h_n/\T)},\hfill
}\right.
\eeq
where the factor
\beq
P_n=\exp(-\lam_n/\T)=\exp(-n/\xidy)
\label{p}
\eeq
describes the {\it a priori}
exponential slowing down of the dynamics with depth $n$.

Throughout the following, $\eps$, $\xidy$, and $\T$
(the latter being referred to as `temperature')
will be considered as three independent parameters of the model.
In particular,
$\eps$ will not necessarily be restricted to the range $0<\eps<1$,
suggested by the interpretation of $(1-\eps$) as the trapped void space.

The zero-temperature statics and dynamics
will crucially depend on whether~$\eps$ is rational or irrational.
Arguing that rational and irrational $\eps$
corresponds to smooth/regular and rough/irregular
shapes, respectively, this difference is to be expected,
and will be discussed further.

\section{Zero-temperature statics}

As the dynamical rules~(\ref{rates}) are fully directional,
they clearly cannot obey detailed balance.
The dynamics simplifies, however, in the $\T\to0$ limit~\cite{usproc},
where~(\ref{rates}) yields
\beq
\frac{w_n(-\to+)}{w_n(+\to-)}=\exp(2\h_n/\T)\to\left\{\matrix{
\infty\hfill&\hbox{if}\hfill&\h_n>0,\cr
0\hfill&\hbox{if}\hfill&\h_n<0.
}\right.
\label{zero}
\eeq
From a purely {\it static} viewpoint,
{\it ground states} of the system can therefore be defined by the condition
that the orientation of every grain is aligned along its local field,
according to the deterministic equation:
\beq
\s_n=\sign\h_n=\left\{\matrix{
+\hfill&\hbox{if}\hfill&\h_n>0,\cr
-\hfill&\hbox{if}\hfill&\h_n<0,
}\right.
\label{zerost}
\eeq
provided $h_n\ne0$~(see below).
The condition~(\ref{zerost})
only involves the parameter $\eps$~[see~(\ref{ydef})].
It is recursive, because of directionality,
in that the right-hand side at depth $n$
only involves the upper grains $k=1,\dots,n-1$.
The uppermost orientation $\s_1$ is left unspecified,
as the corresponding local field vanishes identically.
In the following, we assume for definiteness
that the uppermost grain is ordered:
\beq
\s_1=+.
\label{init}
\eeq

It turns out that the zero-temperature rule~(\ref{zerost})
yields a rich ground-state structure,
because of subtle commensurability and frustration effects.
Our starting point is to observe that~(\ref{zerost}) implies
\beq
\left\{\matrix{
\h_n>0\lra\s_n=+,\quad\hfill& m^+_{n+1}=m^+_n+1,\hfill& m^-_{n+1}=m^-_n,
\hfill&\h_{n+1}=\h_n-1,\hfill\cr
\h_n<0\lra\s_n=-,\hfill& m^+_{n+1}=m^+_n,\hfill& m^-_{n+1}=m^-_n+1,\quad\hfill&
\h_{n+1}=\h_n+\eps.\hfill
}\right.
\label{step}
\eeq
The number and the nature of ground states depend
on whether $\eps$ is rational or irrational, which we consider separately below.

\subsection{Irrational $\eps$: unique quasiperiodic ground state}

Irrational values of $\eps$ imply, in qualitative terms,
a strong irregularity of grain shape.
We refer to this as `roughness'.
Below, we demonstrate that the ground state of a packing of rough grains
is unique, and optimally, but not maximally packed.
We can visualise this as the interlocking
of jutting edges to minimise, but not eliminate, voids.

For irrational $\eps$,~(\ref{step}) implies recursively that
all the local fields $\h_n$ are non-zero, and that they lie in the bounded
interval
\beq
-1\le\h_n\le\eps.
\label{strip}
\eeq

\begin{figure}[htb]
\begin{center}
\includegraphics[angle=90,width=.6\linewidth]{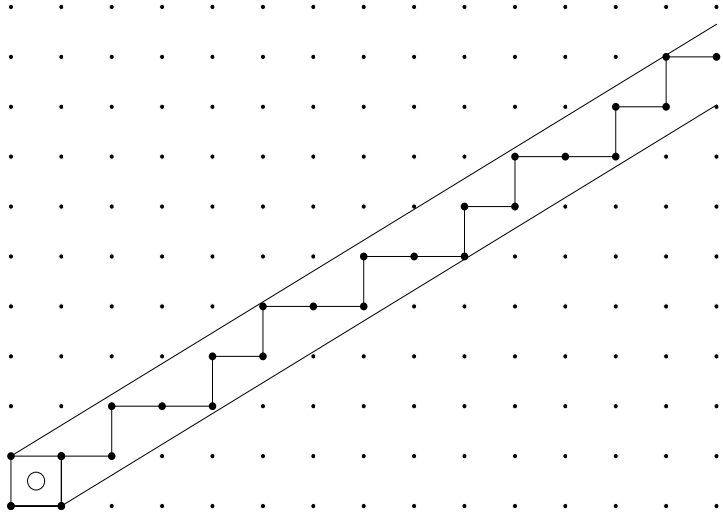}
\caption{\small
Geometrical construction of the quasiperiodic ground state of the model
for the golden-mean slope~(\ref{gold}).
The two ways of going around the first cell, marked with a circle,
correspond to the two possible choices for the orientation
of the uppermost grain.}
\label{figa}
\end{center}
\end{figure}

Let us introduce the following {\it superspace formalism}.
Consider the integers $(m^-_n,m^+_n)$ as the co-ordinates
of points on a square lattice.
We thus obtain a broken, staircase-shaped line, starting as
$(m^-_1,m^+_1)=(0,0)$, $(m^-_2,m^+_2)=(0,1)$ [see~(\ref{init})], etc.
Vertical steps correspond to ordered ($+$) grains,
whereas horizontal steps correspond to disordered ($-$) grains.
Equation~(\ref{strip}) defines an oblique strip with {\it slope} $\eps$
in the $(m^-,m^+)$ plane,
which contains the entire broken line thus constructed~(see Figure~\ref{figa}).

A unique infinite configuration of grain orientations
(i.e., a unique broken line) is thus generated.
This configuration is {\it quasiperiodic}.
Indeed the above construction is equivalent to the cut-and-project method
of generating quasiperiodic tilings of the line,
which has been extensively studied~\cite{quasi}
in the framework of quasicrystals.
(Had we made the initial choice $\s_1=-$ instead of~(\ref{init}),
we would have obtained the same quasiperiodic configuration,
up to a permutation of the two uppermost grains.)
We mention for further reference the following explicit
expressions\footnote{$\Int(x)$, the integer part of a real number $x$,
is the largest integer less than or equal to $x$,
and $\Frac(x)=x-\Int(x)$ is the fractional part of $x$ $(0\le\Frac(x)<1)$.}
for $m_n^\pm$ and $\h_n$:
\beq
m^+_n=n-m^-_n=1+\Int((n-1)\Omega),\qquad
\h_n=-1+\frac{\Frac((n-1)\Omega)}{1-\Omega},
\label{rot}
\eeq
where the rotation number $\Omega$ reads
\beq
\Omega=\eps/(1+\eps).
\eeq
An immediate consequence of~(\ref{rot})
is that there are well-defined proportions of ordered and disordered grains
in the ground state:
\beq
f_+=\Omega=\eps/(1+\eps),\qquad f_-=1-\Omega=1/(1+\eps).
\label{propor}
\eeq

This geometrical construction is illustrated in Figure~\ref{figa}
for the most familiar irrational number, the inverse golden mean~\cite{hr}:
\beq
\eps=\Phi-1=1/\Phi,\qquad\Omega=2-\Phi=1/\Phi^2,\qquad
\Phi=(\sqrt{5}+1)/2\approx1.618033.
\label{gold}
\eeq
The corresponding grain configuration is given by a Fibonacci
sequence~\cite{quasi,hr}:
\[
\{\s_n\}=+--+--+-+--+--+-+--+-+--\cdots
\]

To recapitulate, at any point in the ground state,
the ordering field $h_n$ will assume a finite, non-zero value,
implying in its turn a given orientation of the grain that follows.
The process is therefore fully deterministic, and the state obtained, unique.
We will see that things will be rather different for rational values of $\eps$.

\subsection{Rational $\eps$: degenerate ground states}

Rational values of $\eps$ imply, in qualitative terms,
a regularity of grain shape or of void space, which we term `smoothness'.
We might here expect that regularly shaped grains could align themselves
to fit into accumulated void space, in a given ground state
configuration; this in fact happens,
leading to states of perfect packing at various points of the column.
This in turn gives rise to the observed degeneracy of ground states
to be discussed further below.

For a rational $\eps$:
\beq
\eps=p/q,\qquad\Omega=p/(p+q),
\eeq
in irreducible form ($p$ and $q$ are mutual primes),
some of the local fields $\h_n$ generated by the recursion
equations~(\ref{step}) vanish.
The corresponding grain orientations $\s_n$ remain unspecified.
This means that grain~$n$ has a perfectly packed column above it,
so that it is free to choose its orientation.
For $\eps=1/2$, for example, one can visualise
that each disordered grain `carries' a void half its
size, so that units of perfect packing must be permutations
of the triad $+--$, where the two `half' voids from each
of the ($-$) grains are filled by the ($+$) grain.
(Evidently this is a one-dimensional interpretation
of packing, so that the serial existence of two half voids
and a grain should be interpreted as the insertion of a grain
into a full void in higher dimensions.)
The dynamics, which is {\it stepwise compacting}, selects only two of these
patterns, $+--$ and $-+-$~(see the second line of Table~1).

\begin{figure}[htb]
\begin{center}
\includegraphics[angle=90,width=.6\linewidth]{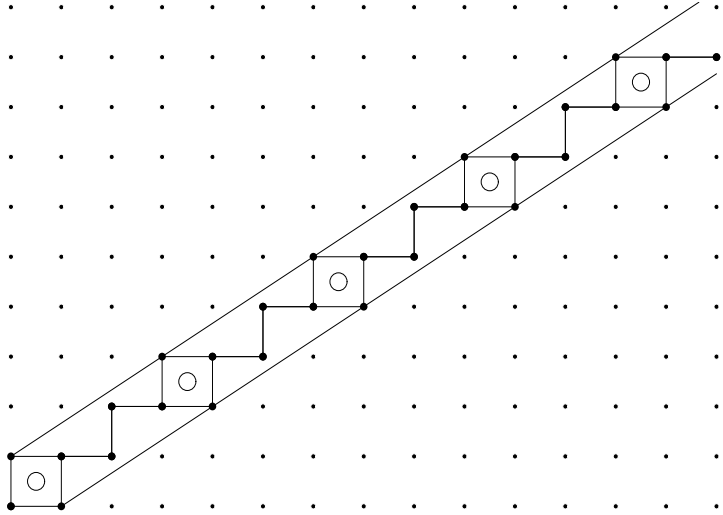}
\caption{\small
Geometrical construction of the ground states of the model
for the rational slope $\eps=2/3$.
The marked cells, entirely contained in the strip,
are responsible for the non-zero configurational entropy.}
\label{figb}
\end{center}
\end{figure}

This feature of rational slopes is clearly visible
on the geometrical construction.
Figure~\ref{figb}, corresponding to $\eps=2/3$,
shows that some of the lattice cells, marked with circles,
are entirely contained in the closed strip~(\ref{strip}).
Consider one such cell.
The broken line enters the cell at its lower left corner
and exits the cell at its upper right corner.
It can go either counterclockwise, via the lower right corner,
giving $\s_{n+1}=-$, $\s_{n+2}=+$,
or clockwise, via the upper left corner, giving $\s_{n+1}=+$, $\s_{n+2}=-$.
Each marked cell thus generates a binary choice in the construction.
This orientational indeterminacy occurs at points of perfect packing,
such that $n$ is a multiple of the {\it period} $p+q$,
equal to the denominator of the rotation number $\Omega$.
The model therefore has a non-zero ground-state entropy,
or zero-temperature configurational entropy, $\Sigma=\ln 2/(p+q)$ per grain.
Each ground state is a random sequence of two
well-defined patterns of length $p+q$,
each of them made of $p$ ordered and $q$ disordered ones,
so that~(\ref{propor}) still holds for each of the ground states.
The patterns only differ by their first two orientations.
The first cases are listed in Table~\ref{tab}.
Finally, the period $p+q$ is formally infinite for an irrational slope.
Accordingly, there is only one marked cell in Figure~\ref{figa}, for $n=0$,
corresponding to the fact that only the uppermost grain is unspecified.

\begin{table}[htb]
\begin{center}
\begin{tabular}{|c|c|c|c|c|c|c|}
\hline
period $p+q$&rot.~number $\Omega$
&slope $\eps$&$p$&$q$&pattern 1&pattern 2\\
\hline
2&1/2&1&1&1&$+-{}$&$-+{}$\\
\hline
3&1/3&1/2&1&2&$+--$&$-+-$\\
3&2/3&2&2&1&$+-+$&$-++$\\
\hline
4&1/4&1/3&1&3&$+---{}$&$-+--{}$\\
4&3/4&3&3&1&$+-++{}$&$-+++{}$\\
\hline
5&1/5&1/4&1&4&$+----$&$-+---$\\
5&2/5&2/3&2&3&$+--+-$&$-+-+-$\\
5&3/5&3/2&3&2&$+-+-+$&$-++-+$\\
5&4/5&4&4&1&$+-+++$&$-++++$\\
\hline
6&1/6&1/5&1&5&$+-----{}$&$-+----{}$\\
6&5/6&5&5&1&$+-++++{}$&$-+++++{}$\\
\hline
\end{tabular}
\caption{\small Patterns building up the random ground states
for the first rational values of $\eps$.
The second example with period 5 is illustrated in Figure~\ref{figb}.}
\label{tab}
\end{center}
\end{table}

\section{Zero-temperature dynamics}

Zero-temperature dynamics is a priori
the canonical way of retrieving the ground states of a system.
Here, the rule for zero-temperature dynamics is:
\beq
\s_n\to\sign\h_n,
\label{zerody}
\eeq
according to~(\ref{zero}), with the definition~(\ref{ydef}).
We will find that irregular grains are able to retrieve
their unique ground state, but that the degeneracy of the ground states
for regularly shaped grains will make them impossible to retrieve.
In the latter case, we find instead a steady state
with non-trivial {\it density fluctuations} above the ground states,
which recall the observed density fluctuations
above the random close-packed state~\cite{sid, gcbam}.

\subsection{Irrational $\eps$, infinite $\xidy$: ballistic coarsening}

For irrational $\eps$, the rule~(\ref{zerody}) is always well-defined,
as the local fields $\h_n$ never vanish.
We start with the situation where $\xidy$ is infinite.
We assume that the system is initially in a disordered state,
where each grain is oriented at random: $\s_n=\pm$ with equal probabilities,
except for the uppermost one, which is fixed according to~(\ref{init}).

The zero-temperature dynamics is observed to drive
the system to its quasiperiodic ground state.
This ordering propagates down the system from its top surface,
via {\it ballistic coarsening}.
At time $t$, the grain orientations have converged
to their ground-state values, given by the above geometrical construction,
in an upper layer whose depth is observed to grow linearly with time:
\beq
L(t)\approx Vt,
\label{vt}
\eeq
whereas the rest of the system is still nearly in its disordered initial state.

\begin{figure}[htb]
\begin{center}
\includegraphics[angle=90,width=.6\linewidth]{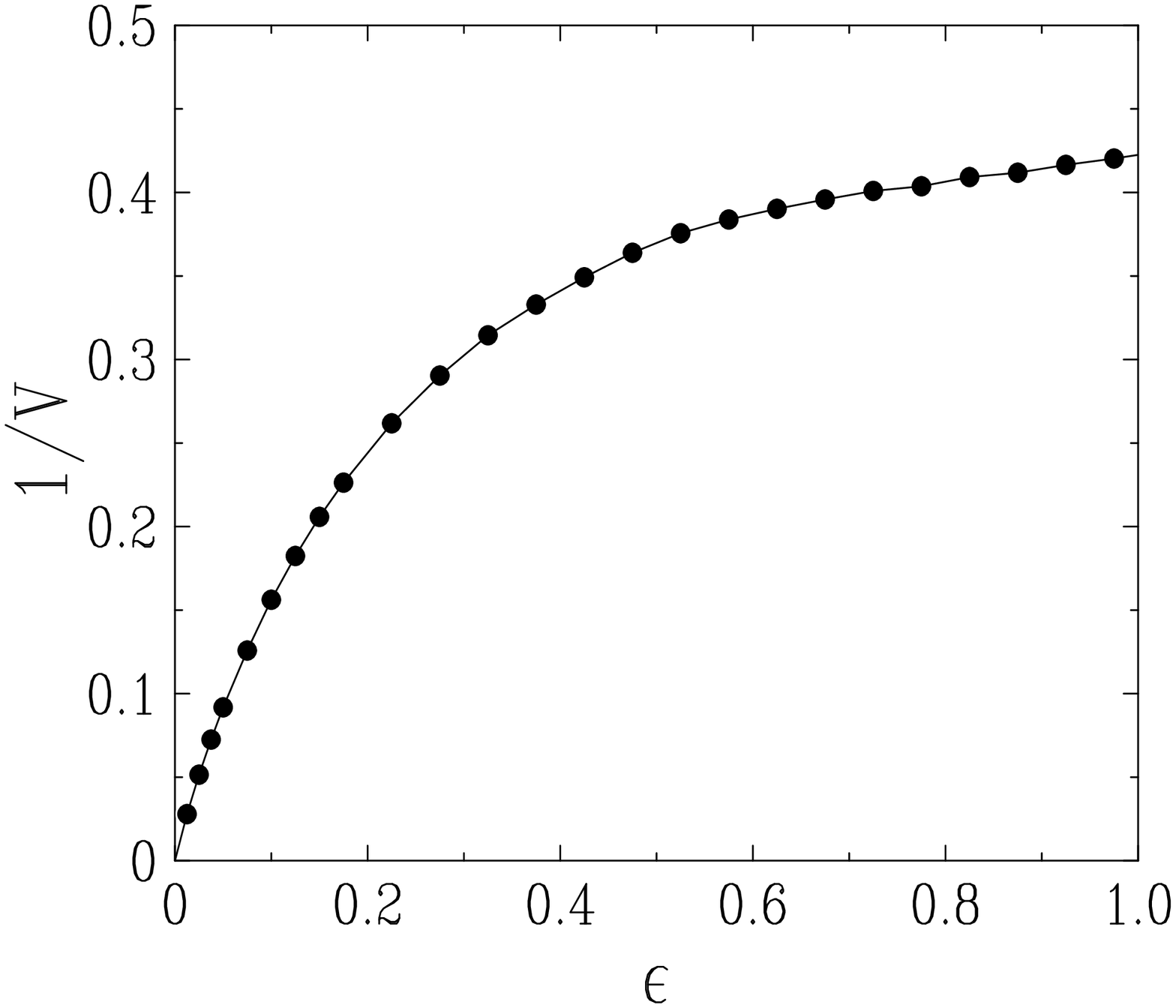}
\caption{\small
Plot of the inverse ordering velocity $1/V$
of zero-temperature coarsening dynamics at infinite $\xidy$,
against the irrational slope $\eps$, for $0<\eps<1$.}
\label{figc}
\end{center}
\end{figure}

This phenomenon is similar to phase ordering,
as order propagates over a macroscopic length $L(t)$ which grows forever.
It is however different from usual coarsening,
as the depth of the ordered region grows ballistically,
vith a well-defined $\eps$-dependent ordering velocity $V$,
instead of diffusively, or even more slowly~\cite{bray}.
Figure~\ref{figc} shows a plot of the inverse of the ordering velocity,
measured in a numerical simulation, against $\eps$, for $0<\eps<1$.
The ordering velocity obeys the symmetry property $V(\eps)=V(1/\eps)$.
It is observed to vary smoothly with $\eps$
(although it is only defined for irrational $\eps$),
and to diverge as $V\sim1/\eps$ as $\eps\to0$.

\subsection{Irrational $\eps$, finite $\xidy$:
crossover to logarithmic coarsening}

For irrational $\eps$, in the situation where $\xidy$ is finite,
but large at the microscopic scale of a grain,
the ballistic coarsening law~(\ref{vt}) is to be modified as
$\d L/\d t\approx V\,\exp(-L/\xidy)$,
taking the slowing down factor~(\ref{p}) into account, hence
\beq
L(t)\approx\xidy\ln(1+Vt/\xidy).
\label{cross}
\eeq

Equation~(\ref{cross}) exhibits a crossover between
the ballistic law~(\ref{vt}) for $1\ll Vt\ll\xidy$,
and the logarithmic coarsening law
\beq
L(t)\approx\xidy\ln t,
\label{log}
\eeq
already present in the model of non-interacting grains~\cite{usepl,usproc}.
The dynamical length $\xidy$ thus controls the spatial dependence
of dynamical behaviour.
In earlier work~\cite{usepl} it was shown to determine the extent to which
order propagates down the column, in the glassy regime.
This interpretation in terms of an {\it ordered boundary layer} continues to be
valid in the present case: For an initially disordered state,
the application of zero-temperature dynamics causes the
quasiperiodic ground state to be recovered
downwards from the free surface to a depth which grows ballistically with time.
When $L(t)$ becomes comparable with $\xidy$,
the effects of the free surface begin to be damped,
and in particular, for $t\gg\xidy/V$, one
recovers the logarithmic law~(\ref{log}), widely associated with
the slow dynamical relaxation of vibrated sand~\cite{sid}.

\begin{figure}[htb]
\begin{center}
\includegraphics[angle=90,width=.6\linewidth]{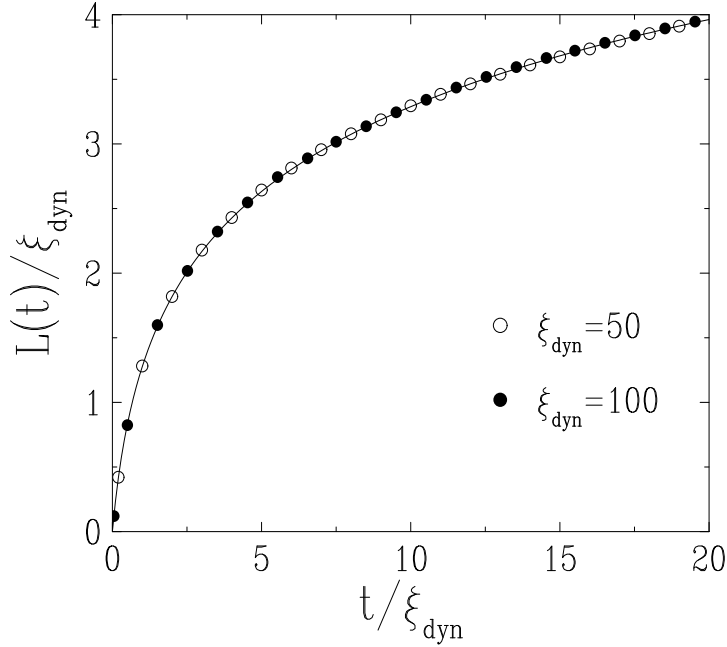}
\caption{\small
Scaling plot of $L(t)/\xidy$ against $t/\xidy$
for zero-temperature coarsening dynamics with the golden-mean slope.
Symbols: numerical data.
Full line: prediction~(\ref{cross}), with $V=2.58$.}
\label{figd}
\end{center}
\end{figure}

Equation~(\ref{cross}) has been checked
against the results of accurate numerical simulations,
for the golden-mean slope.
Figure~\ref{figd} shows a scaling plot of numerical data for $L(t)$
corresponding to $\xidy=50$ and 100, together with the
prediction~(\ref{cross}), with no adjustable parameter.
The ordering velocity $V\approx2.58$
is taken from the data of Figure~\ref{figc}.

\subsection{Rational $\eps$, infinite $\xidy$: anomalous roughening}

We now turn to zero-temperature dynamics for rational $\eps$.
The updating rule~(\ref{zerody}) is not always well-defined as it stands,
as the local fields $\h_n$ may now vanish.
In such a circumstance, it is natural to choose the corresponding orientation
at random:
\beq
\s_n\to\left\{\matrix{
+\hfill&\hbox{if}\hfill&\h_n>0,\cr
\pm\hfill\hbox{ with prob.~}1/2\hfill&\hbox{if}\hfill&\h_n=0,\cr
-\hfill&\hbox{if}\hfill&\h_n<0.\cr
}\right.
\label{zerodyrat}
\eeq
The zero-temperature dynamics defined in this way
therefore keeps a stochastic component.
We focus our attention onto the simplest rational case, i.e., $\eps=1$.
Equation~(\ref{ydef}) for the local fields reads
\beq
\h_n=-\sum_{m=1}^{n-1}\s_m.
\label{hun}
\eeq

We consider first the case where $\xidy$ is infinite.
We observe that the zero-temperature dynamics~(\ref{zerodyrat})
does not drive the system to any of its degenerate dimerised ground states.
The system rather shows a fast relaxation to a unique, non-trivial steady state,
independent of the initial state.
We now investigate this novel kind of zero-temperature steady state
in some detail.

\subsubsection*{Density fluctuations}

\begin{figure}[htb]
\begin{center}
\includegraphics[angle=90,width=.6\linewidth]{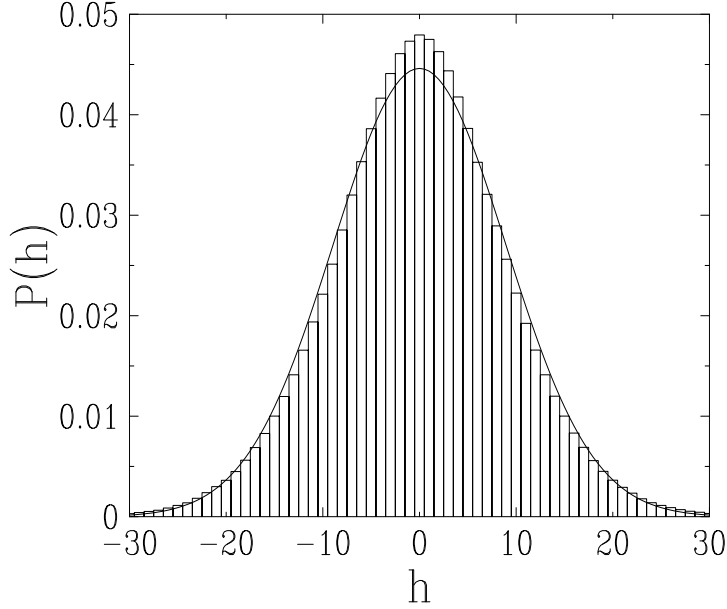}
\caption{\small
Plot of the distribution of the local field $\h_n$ for $n\approx1000$.
Histogram: numerical data (data for $n=999$ and $n=1000$
are mixed in order to avoid spurious parity effects).
Full curve: Gaussian law with width $W_{1000}=8.94$.}
\label{figm}
\end{center}
\end{figure}

\begin{figure}[htb]
\begin{center}
\includegraphics[angle=90,width=.6\linewidth]{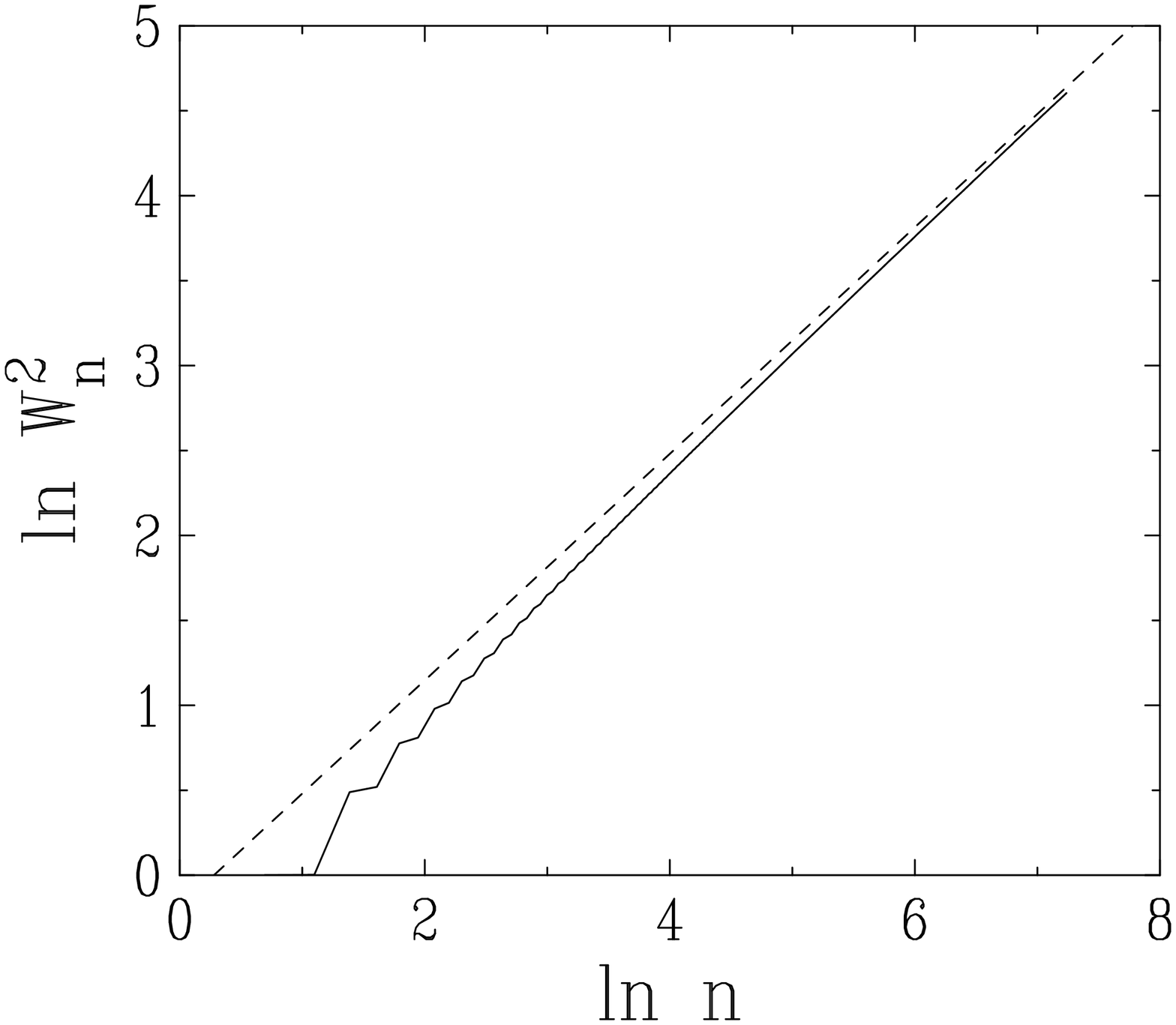}
\caption{\small
Log-log plot of $W_n^2=\mean{h_n^2}$ against depth $n$,
for zero-temperature dynamics with $\eps=1$.
Full line: numerical data.
Dashed line: fit to asymptotic behaviour, leading to~(\ref{rough})
(after~\cite{usletter}).}
\label{fign}
\end{center}
\end{figure}

First of all, the local field $\h_n$ has unbounded fluctuations in the
steady state.
Figure~\ref{figm} shows that these fluctuations
have a Gaussian distribution of width $W_n$,
at least deep enough in the system $(n\gg1)$,
except for a definite excess of small values of the local field:
$\abs{h_n}\sim 1\ll W_n$.
Figure~\ref{fign} (already shown in~\cite{usletter} and reproduced here
for completeness) demonstrates that the local field variance grows as
\beq
W_n^2=\mean{\h_n^2}\approx A\,n^{2/3},
\label{rough}
\eeq
with $A\approx 0.83$.

The exponent $2/3$ of the {\it anomalous roughening} law~(\ref{rough})
can be explained by means of the following local Markovian approximation.
Assume that the local field $\h_n$ obeys an effective
Langevin equation of the form
\beq
\d\h_n/\d t=-a_n\h_n+\eta_n(t),
\eeq
where $\eta_n(t)$ is a white noise so that
$\mean{\eta_n(t)\eta_n(t')}=2D_n\,\delta(t-t')$.
Thus $\h_n(t)$ is a Gaussian variable,
whose steady-state width $W_n$ is given by the Einstein relation:
\beq
W_n^2=\mean{\h_n^2}=D_n/a_n.
\label{steady}
\eeq
The effective parameters $a_n$ and $D_n$ can be estimated as follows.
For the deterministic part,~(\ref{zerodyrat}) implies
\beq
\d\mean{h_n}/\d t=\sum_{m=1}^{n-1}\mean{\s_m-\sign\h_m}
\approx-(1-Q_n)\mean{h_n},
\eeq
where the order parameter $Q_n$ is defined as
\beq
Q_n=\mean{\s_n\sign{\h_n}}.
\label{op}
\eeq
The latter quantity will be shown below to fall off
as $n^{-1/3}$~[see~(\ref{opfall})], implying $a_n\approx1$.
The absence of divergence of the relaxation time $\tau_n=1/a_n$ with $n$
explains the observed fast relaxation to the steady state.
As the fluctuating part is due to the second line of~(\ref{zerodyrat}),
the strength of the noise $D_n$ reads, in some units,
\beq
D_n\approx b\sum_{m=1}^{n-1}\prob\{\h_m=0\}
\approx\frac{b}{\sqrt{2\pi}}\sum_{m=1}^{n-1}\frac{1}{W_m},
\label{diff}
\eeq
assuming that the $\h_n$ have a Gaussian distribution.
Equation~(\ref{steady}) yields
\beq
W_n^2\approx\frac{b}{\sqrt{2\pi}}\sum_{m=1}^{n-1}\frac{1}{W_m},
\eeq
hence the power law~(\ref{rough}), with $A=(9b^2/(8\pi))^{1/3}$.

The anomalous roughening law~(\ref{rough}) for
the fluctuations of the ordering field $h_n$
is the most central feature of the zero-tem\-pe\-ra\-ture steady state
observed for rational~$\eps$.
It implies that, unless arranged by hand,
the known ground states of a system of regularly shaped
objects (i.e., the crystalline ground states) will never be retrieved.
On the contrary, the steady state will be one of {\it density fluctuations}
above the ground state
(which are related to fluctuations of the excess void space).
The present model, to our knowledge, thus contains
the first derivation of a possible source of density
fluctuations in granular media~\cite{sid,gcbam}, which,
here, arise quite naturally from the effect of shape.

These anomalous fluctuations can be put in perspective
with two different physical situations.
First, as already underlined in~\cite{usletter},
the power law~(\ref{rough}) is reminiscent of the domain-growth mechanism
in the low-temperature coarsening regime of the Ising chain
with Kawasaki dynamics~\cite{cks}.
Second, using $n$ to represent time in a random walk,
the above results could be used to explain
the $R_n^2\approx n^{2/3}$ law recently observed
in two-dimensional simulations of the effect of shape
in particulate cages~\cite{kob}.

\subsubsection*{Orientation and local field correlations}

If the grain orientations were statistically independent, i.e., uncorrelated,
one would have the simple result $\mean{h_n^2}=n\eps$,
while~(\ref{rough}) implies
that $\mean{h_n^2}$ grows much more slowly than~$n$.
The orientational displacements of each grain are therefore
{\it fully anticorrelated}.
The anticorrelated orientational displacements are
reminiscent of the {\it bridge collapse} seen in displacement-displacement
correlations of strongly compacting grains~\cite{gcbam};
grain orientational displacements in the direction of vibration
were there seen to be strongly anticorrelated
in jammed regions, as each grain tried to collapse
into the void space trapped by its neighbours.
Interestingly, correlations {\it transverse} to the shaking direction
were~\cite{gcbam} found to be rather small, thus, in
self-consistency terms justifying the choice of a column
model in the present case.
Once again,
if we adopt a kinetic viewpoint and treat $n$ as time in a random walk,
these anticorrelations recall the temporal anticorrelations observed in
recent experiments investigating cage properties
near the colloidal glass transition~\cite{weeks}.

To be more specific,
let us denote the orientation and local field correlation functions~as
\beq
c_{m,n}=\mean{\s_m\s_n},\qquad C_{m,n}=\mean{\h_m\h_n}.
\eeq
Equation~(\ref{hun}) implies
\beq
C_{m,n}=\sum_{k=1}^{m-1}\sum_{\ell=1}^{n-1}c_{k,\ell},\qquad
c_{m,n}=C_{m+1,n+1}-C_{m+1,n}-C_{m,n+1}+C_{m,n},
\label{Cc}
\eeq
and especially $c_{n,n}=C_{n+1,n+1}-2C_{n,n+1}+C_{n,n}=1$,
so that $C_{n,n}-C_{n,n\pm1}\approx 1/2$, and more generally
\beq
C_{n,n}-C_{n,n+k}\approx\abs{k}/2\qquad(\abs{k}\ll n).
\label{cabs}
\eeq

\begin{figure}[htb]
\begin{center}
\includegraphics[angle=90,width=.6\linewidth]{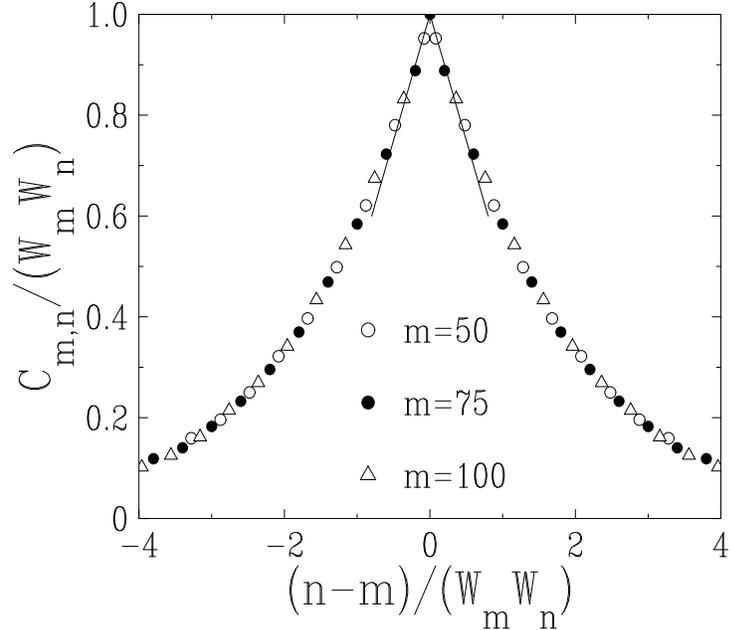}
\caption{\small
Scaling plot of the correlation function $C_{m,n}$
of the local fields in the zero-temperature steady state with $\eps=1$,
demonstrating the validity of~(\ref{Cmn}),
and showing a plot of the scaling function $\F$.
The full lines show the cusp behaviour~(\ref{cusp}).}
\label{figy}
\end{center}
\end{figure}

\begin{figure}[htb]
\begin{center}
\includegraphics[angle=90,width=.6\linewidth]{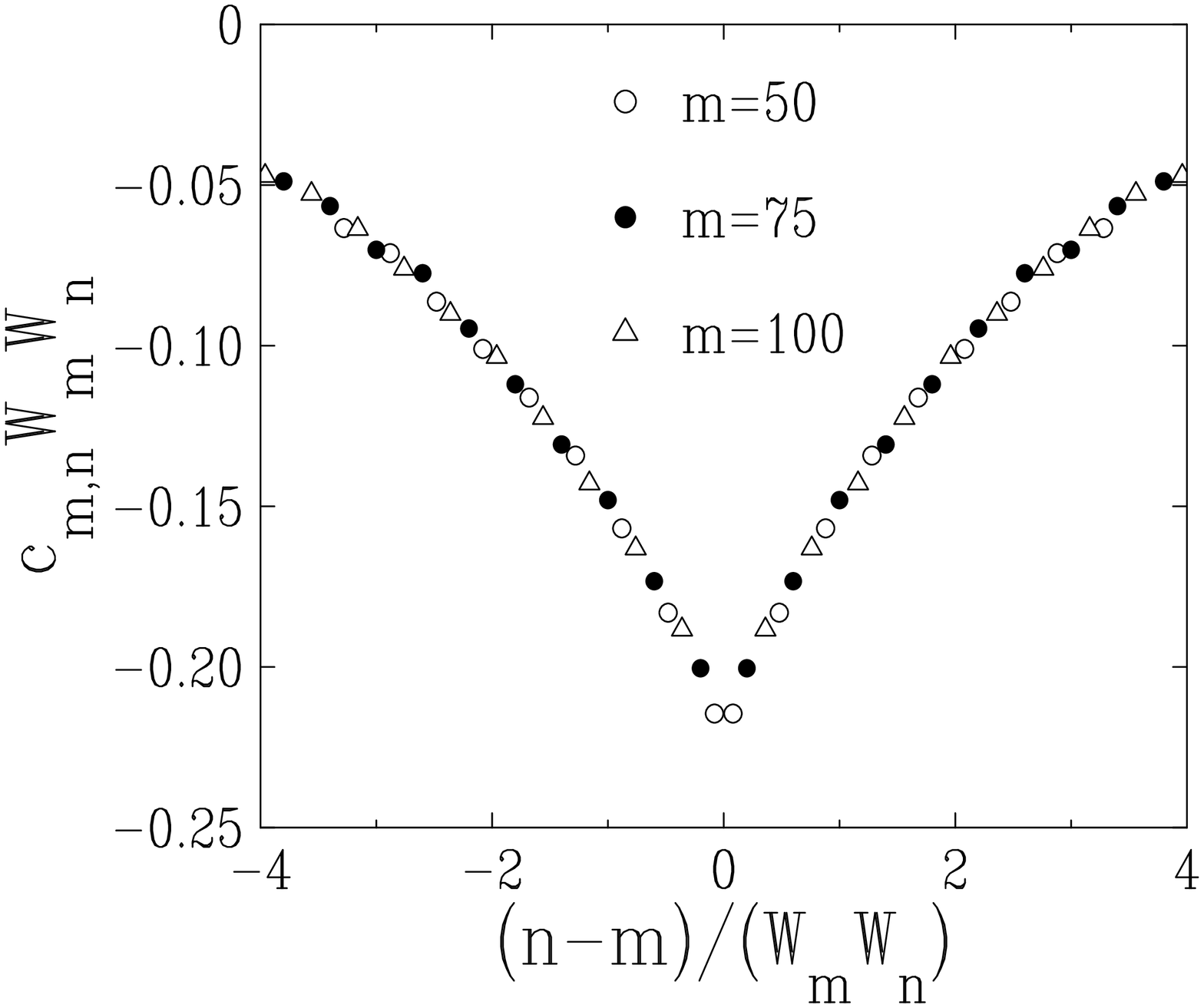}
\caption{\small
Scaling plot of the orientation correlation function $c_{m,n}$ for $n\ne m$
in the zero-temperature steady state with $\eps=1$,
demonstrating the validity of~(\ref{cmn}),
and showing a plot of (minus) the scaling function $F$
(after~\cite{usletter}).}
\label{figo}
\end{center}
\end{figure}

The power law~(\ref{rough}) and the behaviour~(\ref{cabs})
can be combined into the scaling Ansatz
\beq
C_{m,n}\approx W_mW_n\,\F\!\left(\frac{n-m}{W_mW_n}\right),
\label{Cmn}
\eeq
where $\F$ is a positive, even function,
with a cusp at the origin of the form
\beq
\F(x)=1-\abs{x}/2+\cdots\qquad(\abs{x}\ll1).
\label{cusp}
\eeq
As a consequence of~(\ref{Cc}),
the orientation correlations obey a similar scaling law:
\beq
c_{m,n}\approx\delta_{m,n}
-\frac{1}{W_mW_n}\,F\!\left(\frac{n-m}{W_mW_n}\right),
\label{cmn}
\eeq
where $F(x)=\d^2\F/\d x^2$ is another positive, even function such that
\beq
\int_{-\infty}^{+\infty}F(x)\,{\rm d}x=\int_0^{+\infty}x\,F(x)\,{\rm d}x=1.
\eeq
The first of these sum rules confirms
that spin fluctuations are asymptotically {\it totally screened}:
$\sum_{m\ne n}c_{n,n}\approx-c_{n,n}=-1$.
The scaling laws~(\ref{Cmn}) and~(\ref{cmn}) are accurately confirmed
by numerical data for $C_{m,n}$ and $c_{m,n}$,
whose scaling plots are respectively shown
in Figures~\ref{figy} and~\ref{figo}.

\begin{figure}[htb]
\begin{center}
\includegraphics[angle=90,width=.6\linewidth]{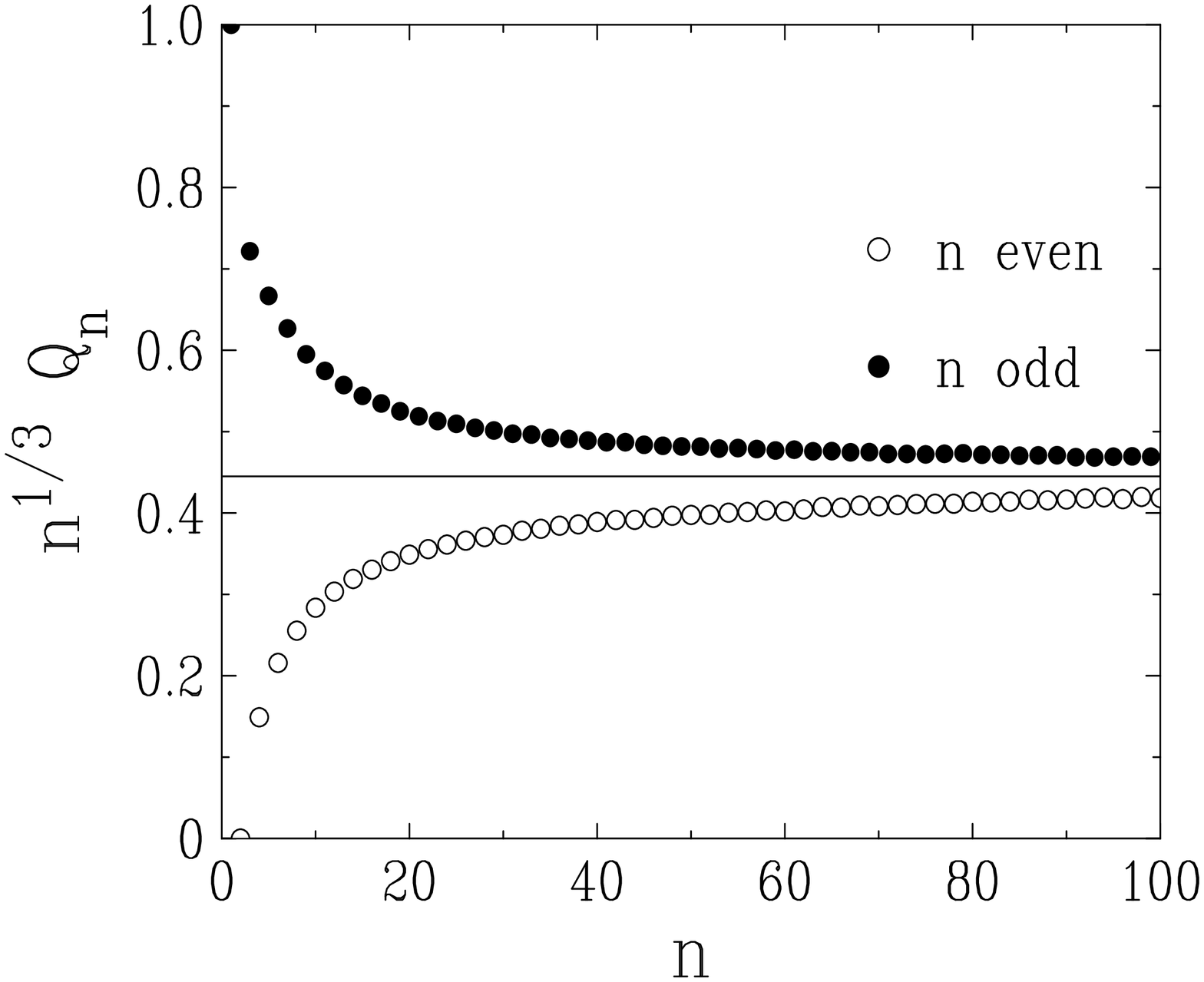}
\caption{\small
Plot of $n^{1/3}$ times the order parameter $Q_n$ against $n$,
for the zero-temperature steady state with $\eps=1$,
Symbols: numerical data.
Full line: common limit value, yielding $a\approx0.44$ in~(\ref{opfall}).}
\label{figg}
\end{center}
\end{figure}

A final consequence concerns the mixed correlation
\beq
\mean{\s_n\h_n}=\sum_{m=1}^{n-1}c_{m,n},
\eeq
for which the scaling results~(\ref{Cmn}),~(\ref{cmn})
yield $\mean{\s_n\h_n}\approx1/2$.
Scaling then implies that the order parameter defined in~(\ref{op})
falls off as $Q_n\sim1/W_n$, hence the estimate
\beq
Q_n\approx a\,n^{-1/3}.
\label{opfall}
\eeq
This power-law decay is well confirmed by numerical data,
shown in Figure~\ref{figg}, which yield $a\approx0.44$.

All of the above adds up, for the case of regularly shaped grains,
to a picture of grain orientations
which are anticorrelated within a {\it dynamical cluster}~\cite{gcbam}
whose size scales as~$n^{2/3}$; grains
{\it outside} such a cluster, are orientationally screened from each other,
i.e., the screening length also goes as $n^{2/3}$.
(Correspondingly, from a kinetic viewpoint~\cite{kob},
these results may be interpreted in terms of the time $n^{2/3}$
spent by a walker bouncing back and forth between the walls of a cage,
where his steps are consequently anticorrelated one with the other.)
Consistently, the order parameter $Q_n$, proportional to the ratio
of the screening length to the total length, goes as
$n^{2/3}/n\sim n^{-1/3}$.
By contrast, when $\eps$ is irrational,
earlier orientations influence {\it all} successive ones,
as the orientation correlations~$c_{mn}$ do not decay to zero.
The order parameter is $Q_n=1$ identically while,
loosely speaking, the screening length scales as $n$.

\subsubsection*{Entropy}

We now turn to the entropy of the steady state,
defined by the usual Boltzmann formula
\beq
S=-\sum_\C p(\C)\ln p(\C),
\label{boltz}
\eeq
where $p(\C)$ is the probability that the system is
in the orientation configuration $\C$ in the steady state,
and the sum runs over all the $2^n$ configurations
$\C=\{\s_m\}$ $(m=1,\dots,n)$ of a system of $n$ grains.

On the theoretical side, the entropy $S$ can be estimated as follows,
using the main feature of the zero-temperature steady state,
i.e., the roughening law~(\ref{rough}).
Think of the depth $n$ as a fictitious discrete time,
and of the local field $\h_n$ as the position of a random walker at time $n$.
For a free lattice random walk of $n$ steps,
one has $\mean{h_n^2}=n$, and the entropy reads $S_\flat=n\ln 2$,
as all configurations are equally probable.
Because $\mean{\h_n^2}=W_n^2\ll n$,
the entropy $S$ of our random walk is reduced with respect to $S_\flat$.
Let
\beq
\Delta S=S_\flat-S=n\ln 2-S
\label{entred}
\eeq
be the entropy reduction~\cite{remi}.
Consider first a strict constraint $\abs{\h_n}<L$.
The probability that a random walk of $n$ steps
obeys this constraint is known to fall off exponentially,
as $\P_n\approx\exp(-\pi^2n/(2L^2))$.
For a slowly time-dependent constraint $\abs{\h_n}<L_n$,
this estimate generalises to
\beq
\P_n\approx\exp\left(-\frac{\pi^2}{2}\sum_{m=1}^n\frac{1}{L_m^2}\right).
\eeq
With the assumption that the strict constraint $\abs{\h_n}<W_n$
and the weak constraint $\mean{\h_n^2}=W_n$
generate similar entropy reductions for similar constraint profiles,
we obtain the estimate
\beq
\Delta S=-\ln\P_n\sim\sum_{m=1}^n\frac{1}{W_m^2}\sim n^{1/3}.
\label{stheo}
\eeq

\begin{figure}[htb]
\begin{center}
\includegraphics[angle=90,width=.6\linewidth]{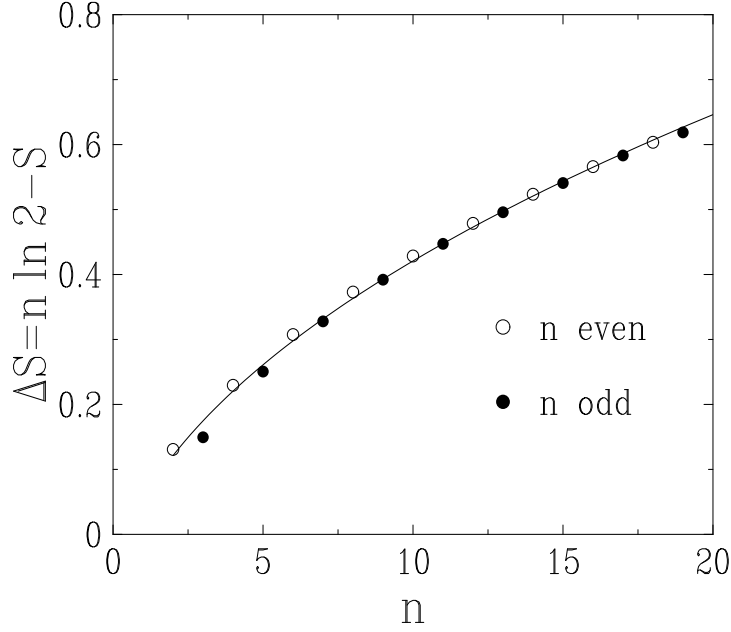}
\caption{\small
Plot of the measured entropy reduction $\Delta S$
in the zero-temperature steady state with $\eps=1$,
defined in~(\ref{entred}), against $n\le19$.
Symbols: numerical data.
Full line: fit $\Delta S=(62\ln n+53)10^{-3}\,n^{1/3}$.}
\label{figh}
\end{center}
\end{figure}

\begin{figure}[htb]
\begin{center}
\includegraphics[angle=90,width=.6\linewidth]{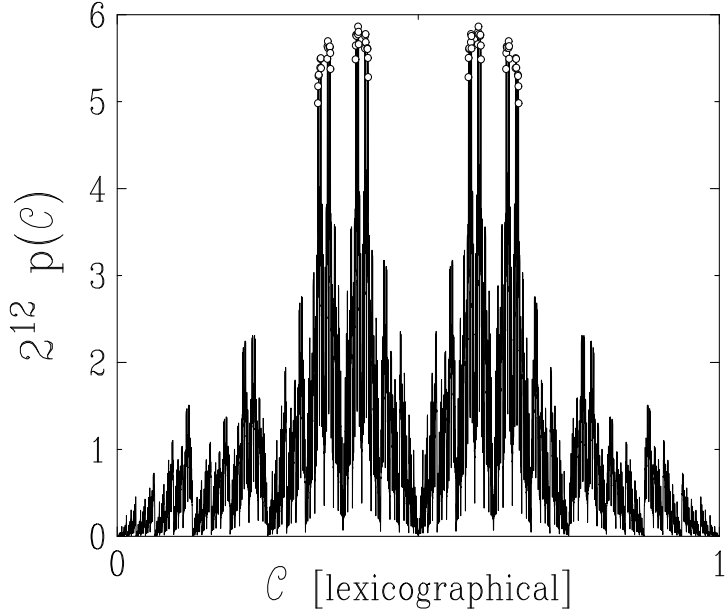}
\caption{\small
Plot of the normalised probabilities $2^{12}\,p(\C)$ of the configurations
of a column of 12 grains in the zero-temperature steady state with $\eps=1$,
against the configurations $\C$ in lexicographical order.
The open circles mark the $2^6=64$ ground-state configurations,
which turn out to be the most probable (after~\cite{usletter}).}
\label{figigs}
\end{center}
\end{figure}

We have evaluated the steady-state entropy $S$
in a numerical simulation, using its definition~(\ref{boltz}),
by measuring the probabilities $p(\C)$ of all the configurations.
As there are $2^n$ configurations for a system of $n$ grains,
the a priori statistical error only decays as $(2^n/t)^{1/2}$.
Reliable data are obtained in this way for $t\sim10^9$ and $n\approx20$.
Figure~\ref{figh} shows a plot of the entropy reduction $\Delta S$ against $n$.
The data show that $\Delta S$ is small,
at least for system sizes reachable by numerical simulations.
For $n=12$ (data of Figure~\ref{figigs}) we have $\Delta S\approx0.479$.
A reasonable semi-quantitative agreement
with the estimate~(\ref{stheo}) is found:
the fit shown in the plot suggests that~(\ref{stheo})
is affected by a logarithmic correction
(which cannot be explained by the simple argument given above),
with a small amplitude around $0.06$.

Figure~\ref{figigs} shows the normalised probabilities $2^{12}\,p(\C)$,
plotted against the $2^{12}=4096$ configurations $\C$
of a column of 12 grains,
sorted according to lexicographical order (read down the column).
This plot exhibits a startlingly rugged structure
on this microscopic scale:
some configurations are clearly visited far more often than others.
It turns out that the most visited configurations
are the $2^6=64$ ground states of the system (empty circles).
We suggest that this behaviour is generic: i.e.,
{\it the dynamics of compaction in the jammed state leads to a microscopic
sampling of configuration space which is highly non-uniform, and reflects
the structure of the ground states}.
In spite of this fine structure, the entropy reduction $\Delta S\sim n^{1/3}$
is subextensive,
and therefore negligible with respect to the free entropy $S_\flat=n\ln 2$,
in qualitative agreement with Edwards' flatness hypothesis~\cite{sam,flatness}.
Our model thus provides a natural reconciliation between
the intuitive perception that not all configurations can be equally visited
in the jammed state where compaction is favoured, and the flatness hypothesis
of Edwards, which is eminently sensible for macroscopically large systems.

\subsection{Rational $\eps$, finite $\xidy$: crossover to Brownian roughening}

In the case where $\xidy$ is finite,
the system still relaxes to a non-trivial steady state,
which is qualitatively similar to that obtained for $\xidy=\infty$,
investigated above.

At the quantitative level, the main effect of the finiteness of $\xidy$
is to induce a nontrivial profile of $W_n^2$.
In the regime where both $n$ and $\xidy$ are large,
the following scaling law is observed
\beq
W_n^2\approx(W_n^2)_\infty\,f(n/\xidy),
\label{xirough}
\eeq
where $(W_n^2)_\infty$ is given by the anomalous
roughening law~(\ref{rough}) of the $\xidy=\infty$ steady state,
which holds more generally for $n\ll\xidy$, so that $f(0)=1$.

\begin{figure}[htb]
\begin{center}
\includegraphics[angle=90,width=.6\linewidth]{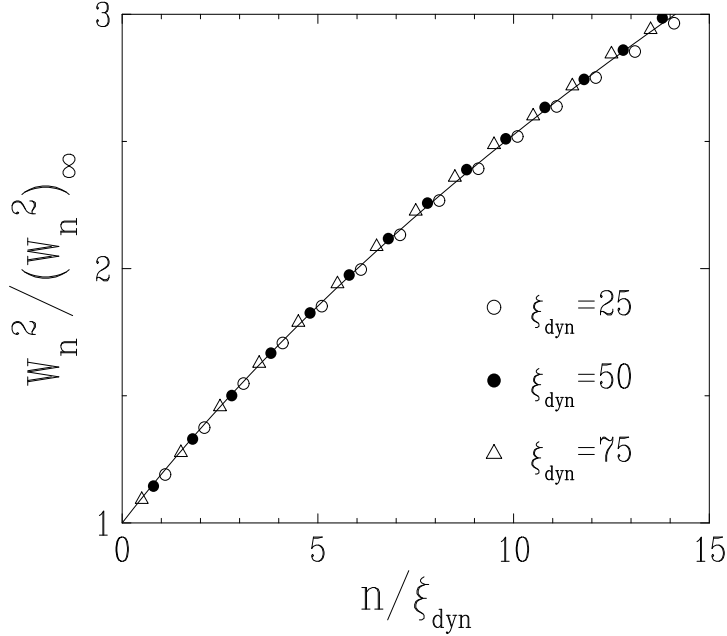}
\caption{\small
Scaling plot of the ratio $W_n^2/(W_n^2)_\infty$ against $x=n/\xidy$
in the zero-temperature steady state with $\eps=1$,
illustrating the scaling law~(\ref{xirough}),
and showing a plot of the scaling function~$f$.
Symbols: numerical data.
Curve: fit $f=1+a((1+bx)^{1/3}-1)$, with $a=5.63$, $b=0.105$,
so that $K=a\,b^{1/3}=2.66$.}
\label{figj}
\end{center}
\end{figure}

A qualitative understanding of the scaling function $f$
can be obtained by generalising the above Markovian approximation.
The expression~(\ref{diff}) for the strength of the noise
is readily replaced by
\beq
D_n\approx\frac{b}{\sqrt{2\pi}}\sum_{m=1}^{n-1}\frac{\e^{-m/\xidy}}{W_m}.
\label{diffxi}
\eeq
For the deterministic part,~(\ref{zerodyrat}) implies
\beq
\d\mean{h_n}/\d t=\sum_{m=1}^{n-1}\mean{\s_m-\sign\h_m}\,\e^{-m/\xidy}.
\eeq
The right-hand side is not simply related to $\h_n$ any more,
so that a further level of approximation is needed.
The most straightforward choice reads
\beq
a_n\approx\frac{1}{n}\sum_{m=1}^{n-1}\e^{-m/\xidy}
\approx\frac{\xidy}{n}(1-\e^{-n/\xidy}).
\label{axi}
\eeq
Skipping the derivation, we mention that~(\ref{diffxi}),
(\ref{axi}) imply~(\ref{xirough}), with
\beq
f(x)=\frac{x^{1/3}}{1-\e^{-x}}
\left(\int_0^x(1-\e^{-y})^{1/2}\,y^{-1/2}\,\e^{-y}\,\d y\right)^{2/3}
\approx
\left\{\matrix{1+x/12\hfill&(x\ll1),\cr K\,x^{1/3}\hfill&(x\gg1),}\right.
\label{xisca}
\eeq
and $K=0.87732$.
In view of the crudeness of the above assumptions,~(\ref{xisca}) is only meant
to provide a qualitative description of the scaling function $f$.
Its asymptotic behaviour for $x\gg1$ is, however,
expected to yield the correct dependence
\beq
W_n^2\approx AK\xidy^{-1/3}n
\label{xibrown}
\eeq
of the width $W_n$ on $n$ and $\xidy$.
The profile of local fields is thus predicted to be Brownian for $n\gg\xidy$.
Figure~\ref{figj} shows a scaling plot of numerical data for the ratio
$W_n^2/(W_n^2)_\infty$, against $x=n/\xidy$.
A scaling law of the form~(\ref{xirough}) is clearly observed.
The fitted curve is compatible with~(\ref{xibrown}), with $K\approx2.66$.

\section{Low-temperature dynamics}

We now turn to the investigation of the low-temperature dynamics of the model.
Our main findings are the observation of {\it intermittency} in the position
of the boundary layer; this has recently been observed in experiments
of vibrated granular beds~\cite{eric}.

We consider for simplicity the case of an infinite $\xidy$.
If the slope $\eps$ is irrational, the dynamical rule~(\ref{zerody})
is fully deterministic at zero temperature,
so that a small non-zero temperature is expected to have drastic effects.
To the contrary, for a rational slope $\eps$,
the rule~(\ref{zerodyrat}) is already stochastic at zero temperature,
and indeed no interesting effect appears at a small non-zero temperature.

We therefore focus our attention onto the case of an irrational slope $\eps$.
We recall that the zero-temperature dynamics drives the system
to its unique quasiperiodic ground state,
where each orientation is aligned with its local field,
according to~(\ref{zerost}).
For a low but non-zero temperature $\T$, there will be {\it mistakes},
i.e., orientations $\s_n=-\sign{\h_n}$ not aligned with their local field.
Equation~(\ref{zero}) suggests that the a priori probability
of observing a mistake at site $n$ scales as
\beq
\Pi(n)\approx\exp(-2\abs{h_n}/\T).
\eeq
Hence the sites $n$ such that the local field $\h_n$ is relatively small
in the ground state $(\abs{\h_n}\sim\T\ll1)$
will be nucleation sites for mistakes,
and thus govern the low-temperature dynamics,
in a sense that will become more precise.

The leading nucleation sites can be located as follows.
Equation~(\ref{rot}) shows that the local field $\h_n$ is small
when $n\Omega$ is close to an integer $m$.
The latter turns out to be $m=m^+_n$.
Indeed
\beq
n\Omega=m+\delta\lra\h_n=\delta/(1-\Omega)
\eeq
for $\delta$ small enough $(\Omega-1<\delta<\Omega)$.
The leading nucleation sites therefore correspond to
the rational numbers $m/n$ which are the closest
to the irrational rotation number~$\Omega$.
This is a well-defined problem of Number Theory,
referred to as the Diophantine approximation~\cite{hr}.

\subsection{The golden-mean slope}
\label{lowgold}

Before we tackle the problem in general,
we consider again for definiteness the golden-mean slope~(\ref{gold}).
In this case, we are led to introduce the Fibonacci numbers
$F_k$~\cite{quasi,hr}, defined by the recursion formula
\beq
F_k=F_{k-1}+F_{k-2}\qquad(F_0=0,\quad F_1=1).
\eeq
We have alternatively
\beq
F_k=\frac{\Phi^k-(-\Phi)^{-k}}{\sqrt{5}}.
\label{fibtau}
\eeq
The leading nucleation sites are the Fibonacci sites $n=F_k$.
We have $m=m^+_n=F_{k-2}$, $m^-_n=F_{k-1}$, and
\beq
\h_n=(-)^k\,\Phi^{-(k-1)},
\eeq
so that
\beq
\Pi_k=\Pi(F_k)\sim\exp\left(-\frac{2\Phi}{\sqrt{5}\,\T F_k}\right).
\label{pik}
\eeq

\begin{figure}[htb]
\begin{center}
\includegraphics[angle=90,width=.6\linewidth]{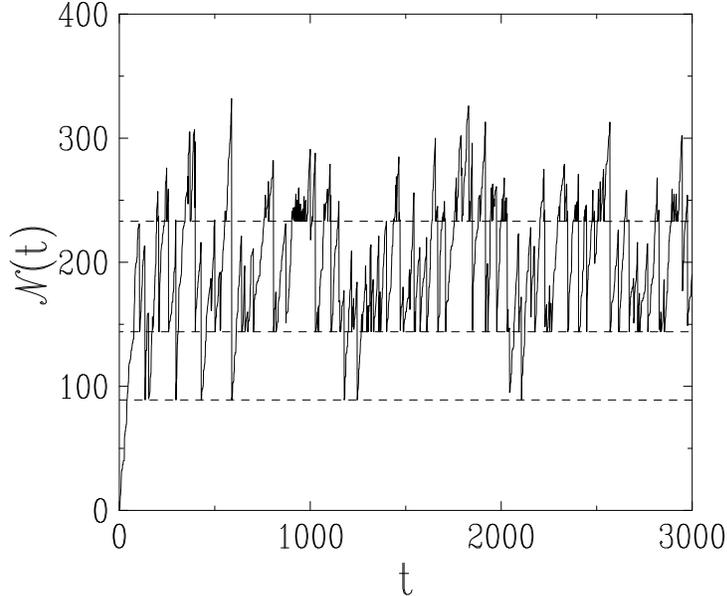}
\caption{\small
Plot of the instantaneous depth $\N(t)$ of the ordered layer,
for the golden-mean slope at $\T=0.003$.
Dashed lines: leading nucleation sites given by consecutive Fibonacci numbers
(bottom to top: $F_{11}=89$, $F_{12}=144$, $F_{13}=233$)
(after~\cite{usletter}).}
\label{figk}
\end{center}
\end{figure}

We can therefore draw the following picture of low-temperature dynamics.
Mistakes are nucleated at Fibonacci sites, according to a Poisson process.
They are then advected with constant velocity $V\approx2.58$,
just as in the zero-temperature case.
The system is ordered according to its quasiperiodic ground state
in an upper layer $(n<\N(t))$, while the rest is disordered,
somehow like the zero-temperature steady state for a rational slope.
The depth $\N(t)$ of the ordered layer,
given by the position of the {\it uppermost} mistake,
is a collective co-ordinate describing low-temperature dynamics.
It evolves according to ballistic advection, i.e.,
$\N(t_1)=\N(t_0)+V(t_1-t_0)$,
until it jumps backward to a smaller depth $\N(t_1)=F_k$,
if another mistake is nucleated there.
Figure~\ref{figk} shows a typical sawtooth plot of the instantaneous
depth $\N(t)$, for a temperature $\T=0.003$.
The leading nucleation sites are observed to be given by Fibonacci numbers.

The system thus reaches a steady state,
characterised by a finite ordering length $\mean{\N}$,
which diverges at low temperature, as mistakes become more and more rare.
The law of this divergence can be predicted by the following argument.
The most active nucleation Fibonacci site is such that
the nucleation time $1/\Pi_k$ is comparable to the advection time
to the next nucleation site $F_{k+1}$,
i.e., $(F_{k+1}-F_k)/V\approx F_k/(\Phi V)$, hence the estimate
\beq
\frac{\Pi_k F_k}{\Phi V}\sim1.
\label{compro}
\eeq
Indeed, less deep sites have too small nucleation rates,
while the mistakes nucleated at deeper sites have little chance to be
the uppermost ones.
Equations~(\ref{pik}) and~(\ref{compro}) yield
\beq
\frac{\sqrt{5}}{2\Phi}\,\T F_k\ln\frac{F_k}{\Phi V}\sim1.
\label{fklnfk}
\eeq
For $\T=0.003$, and for the Fibonacci sites shown in Figure~\ref{figk},
the left-hand side of~(\ref{fklnfk}) respectively reads
$0.56$ for $F_{11}=89$, $1.06$ for $F_{12}=144$, and $1.94$ for $F_{13}=233$.
The estimate~(\ref{fklnfk}) therefore correctly predicts
the observed fact that $F_{12}=144$
is the most active nucleation site at that temperature.

\begin{figure}[htb]
\begin{center}
\includegraphics[angle=90,width=.6\linewidth]{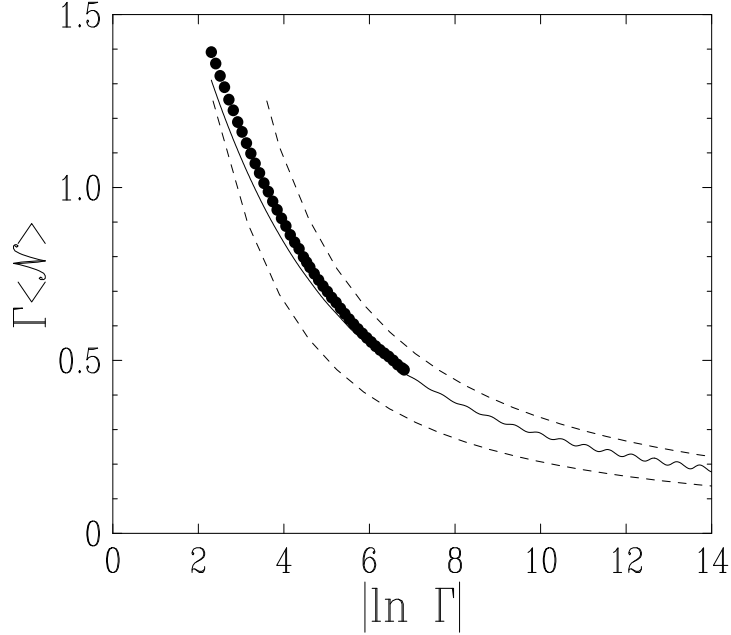}
\caption{\small
Plot of the product $\T\mean{\N}$ against $\absln\T$,
for the golden-mean slope.
Symbols: numerical data.
Full line: analytical prediction~(\ref{nsum}).
Dashed lines: Extrema of the asymptotic result~(\ref{ntimp}),
corresponding to $\A=\A_\max$ (upper curve) and $\A=\A_\min$ (lower curve).}
\label{figp}
\end{center}
\end{figure}

The heuristic argument leading to~(\ref{fklnfk}) can be justified
and made more precise by means of the results of Appendix~A.
The continuum approach used there is justified
because the Fibonacci sites are more and more sparse.
In the case of present interest,
keeping only the Fibonacci sequence of leading nucleation sites,
we obtain the prediction~(\ref{nsum}) for the ordering length,
to be shown in Figure~\ref{figp}.

For a low enough temperature $\T$,
the sum entering the right-hand side of~(\ref{nsum}) is sharply cutoff.
It can indeed be argued that the term of order $k$ in that sum
is essentially $F_{k-1}$ for $k\le k^\star$,
while it is exponentially negligible for $k\ge k^\star+1$,
where $k^\star=\Int(K)$, and $K$ is the {\it real} solution of~(\ref{fklnfk}),
considered as a strict equality,
with $F_K\approx\Phi^K/\sqrt{5}$, according to~(\ref{fibtau}).
We have therefore $\mean{\N}\approx F_{k^\star+1}$, i.e., more explicitly,
\beq
\mean{\N}\approx F_K\,\A_K.
\label{ntimp}
\eeq
The first factor of this expression,
\beq
F_K\approx\frac{2\Phi}{\sqrt{5}\,\T\absln\T}
\left(1-\frac{1}{\absln\T}\,\ln\frac{2\Phi}{\sqrt{5}V\absln\T}+\cdots\right),
\eeq
shows that the ordering length obeys a linear divergence at low temperature,
with explicit logarithmic corrections.
The second factor,
\beq
\A_K=\Phi^{1-\Frac(K)},
\label{oscak}
\eeq
is a periodic function of its argument $K\approx\absln\T/\ln\Phi$,
with unit period,
which oscillates between the bounds $\A_\max=\Phi$ and $\A_\min=1$.
{\it Oscillatory amplitudes} are commonly observed
in models related to self-similar structures~\cite{l};
they originate in the discrete self-similarity of the underlying sequence.
The oscillations of the asymptotic amplitude $\A_K$,
given in~(\ref{oscak}), are damped, except at extremely low temperature.
Figure~\ref{figp} shows a plot of numerical data
for the product $\T\mean{\N}$, against $\absln\T$.
These data are well described by the analytical prediction~(\ref{nsum}),
and lie within the bounds of the asymptotic
estimate~(\ref{ntimp})--(\ref{oscak}).
The oscillations become visible on the analytical curve
for the lower temperatures $(\T<10^{-4})$,
which are not directly accessible to simulations.

\subsection{Other irrational slopes}

We now consider briefly the case of an arbitrary irrational slope $\eps$.
The situation is rather similar to the phenomenon of hierarchical melting,
observed at low temperature in some incommensurate modulated solids~\cite{hime}.

The nucleation sites can be determined as follows.
The irrational rotation number $\Omega$ can be written as an infinite
{\it continued-fraction expansion}~\cite{hr}:
\beq
\Omega=\frac{1}{a_1+\frac{1}{a_2+\frac{1}{a_3+\cdots}}}=[a_1,a_2,a_3,\dots].
\eeq

The principal approximants of $\Omega$ are the rationals
\beq
\Omega_k=p_k/q_k,
\eeq
whose numerators and denominators obey the same linear recursion
\beq
p_k=a_kp_{k-1}+p_{k-2},\qquad q_k=a_kq_{k-1}+q_{k-2},
\eeq
with $p_0=q_{-1}=0$, $p_1=q_0=1$.
The denominators build the leading sequence of nucleation sites.
The rotation number $\Omega$ also has secondary approximants
\beq
\Omega_{k,b}=(bp_{k-1}+p_{k-2})/(bq_{k-1}+q_{k-2})
\eeq
for $b=1,\dots,a_k-1$ if $a_k>1$.
The denominators are the subleading nucleation sites.

For the golden-mean slope, we have $\Omega=2-\Phi=[2,1,1,1,1,\dots]$,
so that the leading Fibonacci nucleation sites of section~\ref{lowgold}
are recovered, whereas there are no subleading nucleation sites.

The most active nucleation site at low temperature
can again be estimated by comparing the nucleation time and the advection time.
The ordering length $\mean{\N}$ is thus still predicted to diverge as
\beq
\mean{\N}\approx\frac{\A(\ln\T)}{\T\absln\T},
\eeq
at least for irrational numbers with typical Diophantine properties.
Most irrational numbers are typical in this respect.
The presence of secondary approximants
makes however the oscillation pattern of the amplitude $\A(\ln\T)$
more complex than a simple periodic function in general,
in analogy with the low-temperature specific heat peaks
induced by hierarchical melting~\cite{hime}.

In more physical terms, the ordering length $\mean{\N}$
defines the mean position of an {\it intermittent} boundary layer,
separating an ordered state above it from a disordered state below.
This length is thus a kind of finite-temperature equivalent of the
`zero-temperature' length $\xidy$.
Both $\mean{\N}$ and $\xidy$ retain the flavour of a boundary layer separating
order from disorder.
Within each of these boundary layers, the relaxation is {\it fast},
and based on single-particle relaxation, i.e., individual particles
attaining their positions of optimal local packing~\cite{gcbam,bergmehta}.
The {\it slow} dynamics of
cooperative relaxation only sets in for lengths {\it beyond} these,
when the moves by which packing needs to be optimised become non-local.

\section{Discussion}

The work we have presented here concerns the effect of
shape on compaction properties of grains.
If irrational and rational values of $\eps$,
the volume occupied by a disordered grain,
are taken to correspond to irregular and regular grains,
we see a distinct difference in behaviour between the two.

While the uniqueness of packing of irregular grains
in their ground state vis-a-vis the degeneracy
of perfectly packed ground states for regularly shaped grains
is in accord with intuition, the effect of dynamics is more subtle.
The fact that the perfectly packed and degenerate
ground states of the regular grains are {\it never} retrieved
by zero-temperature dynamics, leading instead to density
fluctuations (that have been observed experimentally~\cite{sid})
is rather subtle, as compared with the less perfect, unique and perfectly
retrievable ground state for the irregular grains.
Clearly, a sharp distinction between neighbouring rational and irrational
values of $\eps$ only makes sense for an infinitely deep system.
For a finite column made of $N$ grains,
the distinction is rounded off by finite-size effects.
In particular, the characteristic features of any `large' rational $\eps$
are no longer observed when the period $p+q$ becomes larger than $N$.

Zero-temperature steady-state density fluctuations for regular grains
are subextensive: grain orientations are thus {\it fully anticorrelated},
reminiscent of dynamical heterogeneities in bridge collapse~\cite{gcbam} in
strongly compacted granular media, as well as temporal
anticorrelations in cages~\cite{kob,weeks}.
Also, while the macroscopic entropy of the steady state~\cite{remi}
is approximately that of a fully disordered column,
consistent with Edwards' `flatness' hypothesis~\cite{sam},
a microscopic examination of the configurations
reveals a rugged landscape, with the most visited configurations
corresponding to the ground states.
As a matter of fact, the steady-state fluctuations of the local fields
remain subextensive at any finite temperature.
As a consequence, after a short transient regime,
the steady-state proportions of ordered and disordered grains
are solely determined by $\eps$, and given by~(\ref{propor}),
for both rational and irrational values of $\eps$.

Lastly, the low-temperature dynamics for irrational $\eps$
leads to an intermittency of the surface layer.
We would expect that for large enough temperatures,
there would be little distinction between regular and irregular grains;
our model however provides an interesting prediction of boundary-layer
intermittency,
which should be visible at low enough temperatures for irregularly shaped
objects.

\subsubsection*{Acknowledgements}

AM warmly acknowledges the hospitality of the Service de Physique
Th\'eorique, Saclay, where most of this work was conceived.

\newpage
\appendix
\section{Distribution of the depth $\N$ of the upper layer}

This Appendix is devoted to the depth $\N(t)$ of the ordered layer
for low-temperature dynamics in the irrational case.
Our main goal is to derive the stationary distribution of~$\N$,
for the effective dynamics described in section~\ref{lowgold}.

For convenience we use a continuous formalism,
treating $\N$ as a real variable.
Let $\pi(x)\,\d x$ be the given nucleation rate per unit time
between $x$ and $x+\d x$,
and $p(x,t)\,\d x$ be the probability of finding the depth $\N$
between $x$ and $x+\d x$ at time $t$.
The unknown probability distribution function $p(x,t)$ obeys the rate equation
\beq
\left(\frac{\partial}{\partial t}+V\frac{\partial}{\partial x}\right)p(x,t)
=\pi(x)P(x,t)-p(x,t)\Pi(x)
\equiv\frac{\partial}{\partial x}\Bigl(\Pi(x)P(x,t)\Bigr),
\label{dpar}
\eeq
with the notations
\beq
P(x,t)=\int_x^\infty p(y,t)\,\d y,\quad
p(x,t)=-\frac{\partial P(x,t)}{\partial x},\quad
\Pi(x)=\int_0^x\pi(y)\,\d y,\quad
\pi(x)=\frac{\d\Pi(x)}{\d x}.
\eeq
Indeed, the left-hand side of~(\ref{dpar}) is the usual covariant derivative,
whose convective term involves the drift velocity $V$.
The middle side represents the evolution due to nucleation events,
with the first (gain) term originating in nucleation at depth $x$,
and the second (loss) term originating in nucleation at depth $y<x$.

The stationary (time-independent) solution $p_\st(x)$
of~(\ref{dpar}) is such that
\beq
V\,p_\st(x)\equiv-V\,\frac{\d P_\st(x)}{\d x}=\Pi(x)P_\st(x).
\eeq
This separable differential equation easily yields the results
\beq
P_\st(x)=\exp\left(-\frac{1}{V}\int_0^x\Pi(y)\,\d y\right),\quad
p_\st(x)=\frac{\Pi(x)}{V}\exp\left(-\frac{1}{V}\int_0^x\Pi(y)\,\d y\right),
\eeq
and especially
\beq
\mean{\N}
=\int_0^\infty\exp\left(-\frac{1}{V}\int_0^x\Pi(y)\,\d y\right)\d x
=\int_0^\infty\exp\left(-\frac{1}{V}\int_0^x(x-y)\pi(y)\,\d y\right)\d x.
\eeq
These expressions hold for an arbitrary distribution of nucleation rates.

In the case of interest in section~\ref{lowgold},
taking into account the leading sequence of Fibonacci sites $F_k$,
with nucleation rates $\Pi_k$, we obtain
\beq
\mean{\N}=\sum_{k=0}^\infty\frac{V}{A_k}\,
\left(\e^{-B_k/V}-\e^{-B_{k+1}/V}\right),
\label{nsum}
\eeq
with
\beq
A_k=\sum_{\ell=0}^k\Pi_\ell,\quad B_k=\sum_{\ell=0}^{k-1}(F_k-F_\ell)\Pi_\ell.
\eeq

\end{document}